# Exploring the Feasibility of Multimodal Chatbot AI as Copilot in Pathology Diagnostics: Generalist Model's Pitfall

Mianxin Liu [1,†], Jianfeng Wu [2,†], Fang Yan[1], Hongjun Li[1], Wei Wang[3,4], Shaoting Zhang[1,*], Zhe Wang[2,3,4,*]

[1] Shanghai Artificial Intelligence Laboratory, Shanghai, China
[2] State Key Laboratory of Holistic Integrative Management of Gastrointestinal Cancer, Department of Pathology, School of Basic Medicine and Xijing Hospital, Fourth Military Medical University, China
[3] Department of Pathology, The First Affiliated Hospital of University of Science and Technology of China, Division of Life Sciences and Medicine, University of Science and Technology of China, Hefei, Anhui, 230036, China
[4] Intelligent Pathology Institute, Division of Life Sciences and Medicine, University of Science and Technology of China, Hefei, Anhui, 230036, China

†: These authors contributed equally
*: Correspondence to Shaoting Zhang (zhangshaoting@pjlab.org.cn) and Zhe Wang (zhwang@fmmu.edu.cn)

**Abstract**

Pathology images are crucial for diagnosing and managing various diseases by visualizing cellular and tissue-level abnormalities. Recent advancements in artificial intelligence (AI), particularly multimodal models like ChatGPT, have shown promise in transforming medical image analysis through capabilities such as medical vision-language question answering. However, there remains a significant gap in integrating pathology image data with these AI models for clinical applications. This study benchmarks the performance of GPT on pathology images, assessing their diagnostic accuracy and efficiency in real-word clinical records. We observe significant deficits of GPT in bone diseases and a fair-level performance in diseases from other three systems. Despite offering satisfactory abnormality annotations, GPT exhibits consistent disadvantage in terminology accuracy and multimodal integration. Specifically, we demonstrate GPT's failures in interpreting immunohistochemistry results and diagnosing metastatic cancers. This study highlight the weakness of current generalist GPT model and contribute to the integration of pathology and advanced AI.

## Introduction

The field of medical imaging has seen remarkable advancements, with high-resolution imaging playing a critical role in diagnosing and managing various diseases. Pathology images, particularly those obtained from hematoxylin and eosin (H&E) staining, are pivotal in providing detailed insights into the cellular and tissue-level abnormalities associated with various conditions [1, 2]. These images are essential for accurate diagnosis, prognosis, and treatment planning in pathology.

Simultaneously, artificial intelligence (AI) models are trending the digital pathology analyses [3, 4, 5]. And the emergence of multimodal AI models, such as ChatGPT-4V, could bring more transformative changes to pathology image analysis [6, 7]. These advanced AI systems demonstrate the ability to perform medical vision-language question answering, where the AI comprehends the content of pathology images and provides diagnostic assessments interactively [8, 9]. Additionally, these multimodal AI models show promise for automated report generation, enhancing the efficiency and accuracy of medical imaging workflows [10]. Such AI technologies could streamline the diagnostic process and contribute to personalized medicine services.

Despite these technological advancements, a notable gap exists in integrating pathology image data with multimodal AI models for clinical applications. Given the complexity and variability of pathology images, it is uncertain whether these AI models can fully leverage the detailed visual information provided by such images [11, 12]. Therefore, it is crucial to investigate the performance of multimodal AI on pathology image data. However, a comprehensive benchmark for evaluating multimodal AI model in pathology in clinical setting remains lacked. The existing benchmark datasets for this endeavor, such as PathVQA [13] and Quilt-VQA [14] are constructed on noisy internet data and does not ensure the quality. Also, during the construction and the evaluation of existing benchmarks, an absent of supervisions from the pathology experts raises serious concern about the reliability of the results.

This paper aims to benchmark the capabilities of GPT-based model on pathology images. We establish a new benchmark using multimodal data of cases from clinics, covering common and rare diseases, and perform evaluations with inputs from pathology experts. A multi-faceted and case-by-case scoring and commenting reveals insights about the lack of generalist GPT model in pathological applications. By exploring the potential of GPT in clinical settings, we seek to bridge the gap between prevalent pathological imaging technology and cutting-edge AI applications, ultimately contributing to improved clinical efficiency, patient care, and personalized medicine.

## Methods

*Pathology VQA benchmark dataset construction*

To establish the pathology VQA benchmark dataset, we utilized the collected relevant cases in the diagnostic database of the Pathology Department of Xijing Hospital, from January 2023 to May 2024. The cases were primarily collected from the following four systems: bone, ovary, central nervous system (CNS), and liver. For each system, the included cases mainly involve common diseases occurring in these systems, with 1-2 difficult and rare cases selected as well. A total of 39 cases were collected: 10 cases each from the bone, ovary, and CNS, and 9 cases from the liver. For each case, the multiple aspects of information are involved, including age, gender, sample location, medical history, macroscopic imaging data (magnetic resonance, computed tomography, or X-ray imaging results if available), diagnostic results, 1-2 H&E images of typical lesions with a magnification of 200x, and immunohistochemistry (IHC) results (if available). Based on these data, we design 62 rounds of question answering (there can be more than one question for each cases).

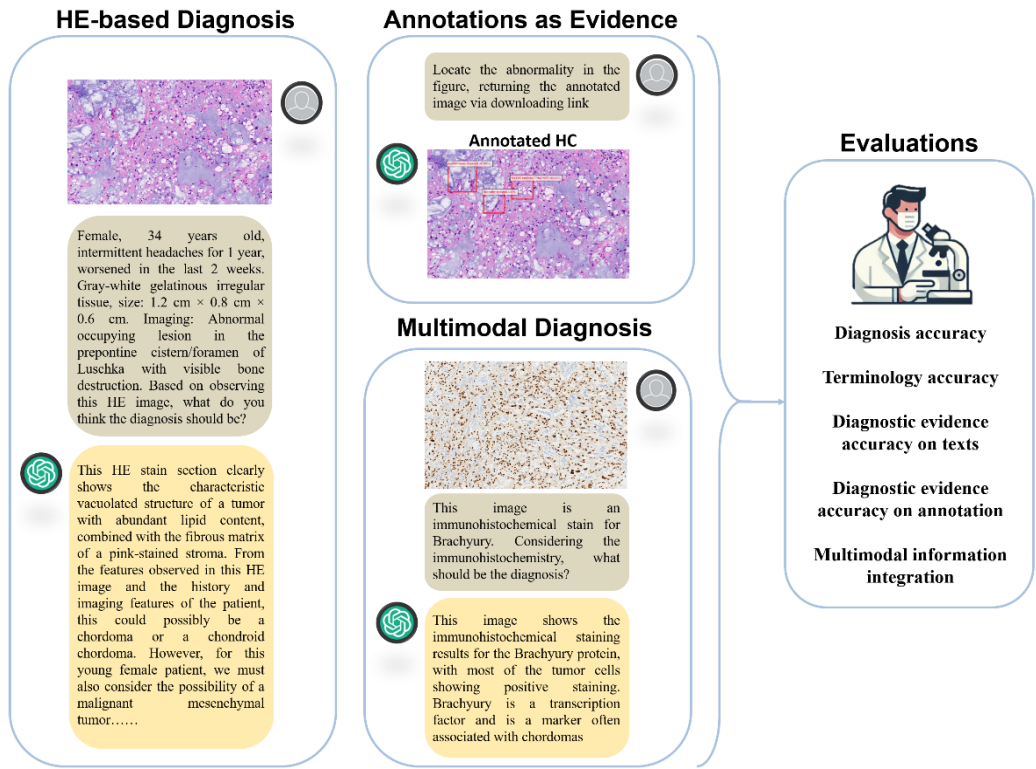

**Figure 1**. The overall workflow of the benchmark process. We prompt GPT to perform H&E-based diagnosis, annotating the image to provide evidence, and multimodal diagnosis. The responses are collected and evaluated by pathologists. The evaluation dimensions includes diagnosis accuracy, terminology accuracy, diagnosis evidence accuracy on texts, diagnosis evidence accuracy on annotations, multimodal information integration.

*Experimental design*

We adopted ChatGPT-4V (https://chat.openai.com/) as the representative multimodal AI model to be tested with the prepared data. We manually selected the most typical slices containing the tumor's vision feature and converted them as 2D PNG images, which will be combined with different pre-defined questions and inputted into ChatGPT-4V. The conversation with GPT was conducted using Chinese and we present the English translations in the result part.

The workflow of the benchmark process is depicted in Fig. 1. We simulated an image reading and diagnosis process HE or IHC images. ChatGPT-4V was instructed to response to the given question based on the presented H&E or IHC images and make diagnosis. We also required GPT to identify the observable abnormality as the evidence supporting the diagnosis, by adding a prompt as "Make textual response and locate the abnormality in the figure, returning the annotated image via downloading link". In part of the questions, we examined whether ChatGPT-4V is capable of integrating information of H&E and IHC images to give accuracy diagnosis.

*Statistical analysis*

A human-based evaluation was implemented to the evaluate the textual answers and thus the quantifications can involve more aspects than barely accuracy. Three experienced experts were instructed to read the image, question, and the corresponding answers from ChatGPT-4V and to rate the answer from 1 to 5. Scores were given from aspects of diagnosis accuracy, terminology accuracy, diagnostic evidence accuracy on texts and annotations, and multimodal information integration capability. We implement a joint review and score process, where the results are firstly discussed among the three experts and only the final results are presented.

# Results

## *Quantitative analysis on the GPT responses*

We designed 62 rounds of question answering based on the real-word clinical data from 39 cases (ovary: 16 rounds, bone: 17 rounds, CNS: 16 rounds, and liver: 13 rounds). The responses from GPT were evaluated by three senior pathology experts. For each question answering, a final score was achieved after a joint review and discussion. Below we presented the results for each system.

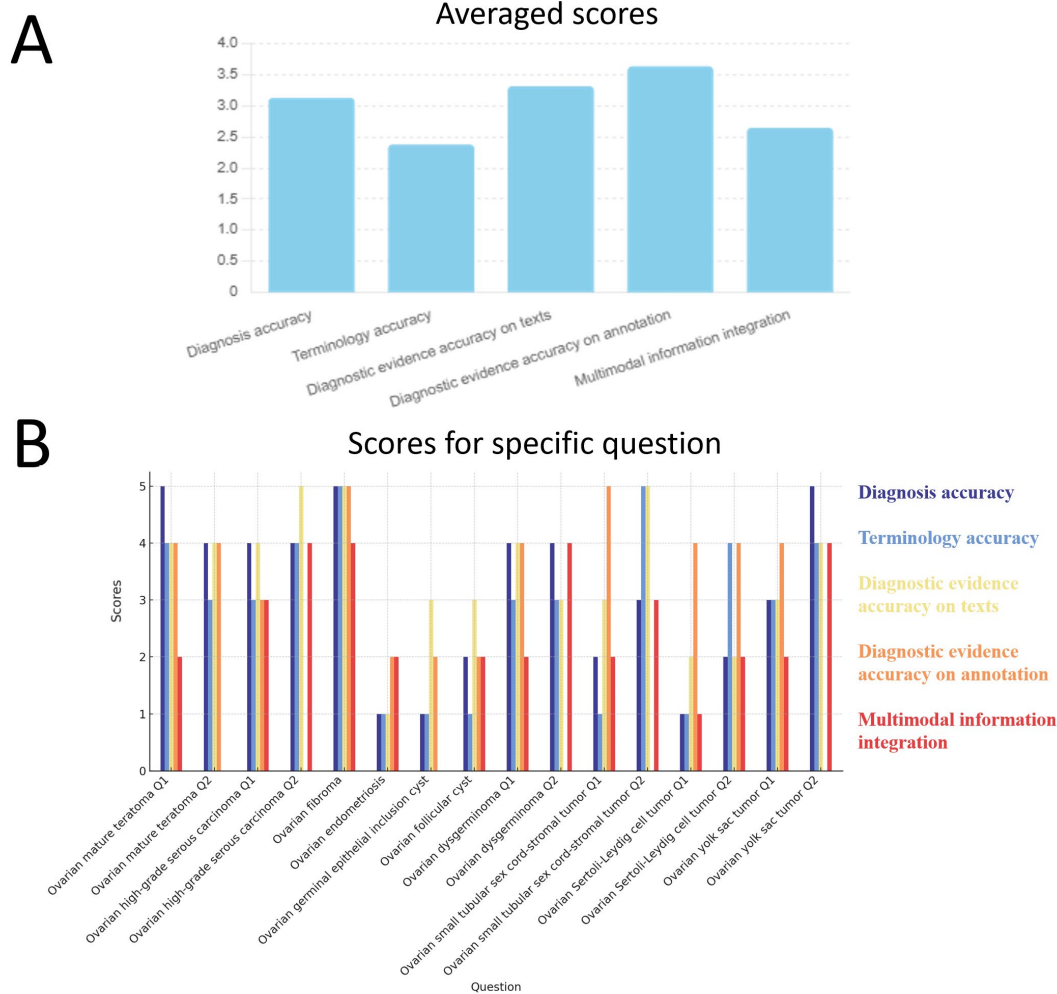

**Figure 2**. The detailed and averaged scores on the response of GPT for questions about ovary system. The original scores range from 1 to 5 and a zero-score indicates "non-applicable".

**Ovary:** In evaluating ChatGPTs performance on ovarian pathology tasks, we observed variable results across different metrics. The overall scores (Fig. 2A) indicate that ChatGPT is satisfactory in diagnostic annotation (average score of 3.64) and diagnosis accuracy (3.13) but shows room for improvement in terminology accuracy (2.38) and multimodal information integration (2.64). Detailed question-level analysis (Fig. 2B) depicts similar performance in cases. The best case is "Ovarian fibroma" while the worst case is "Ovarian endometriosis". Overall deficits can be found in terminology use and integrating multimodal information in most of the cases.

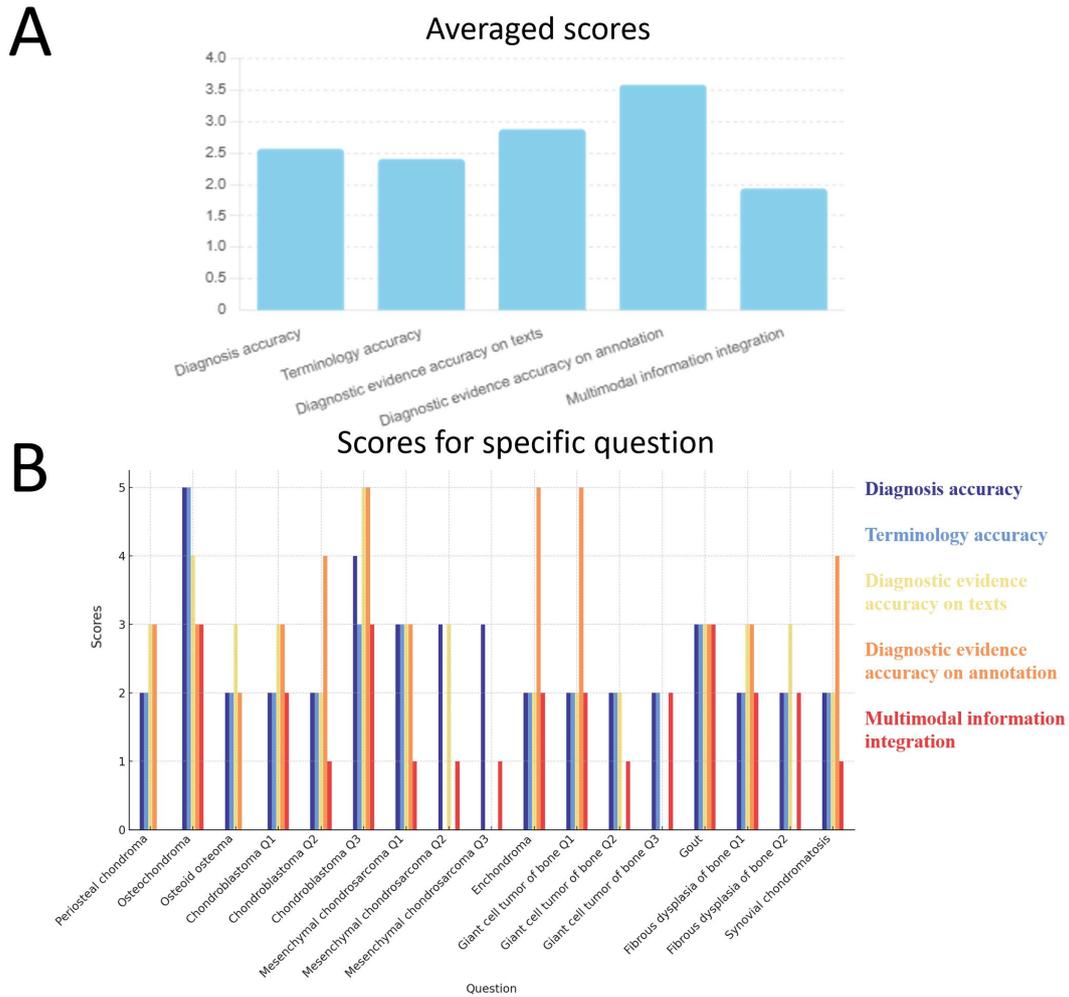

**Figure 3**. The detailed and averaged scores on the response of GPT for questions about bone system. The original scores range from 1 to 5 and a zero-score indicates "non-applicable".

**Bone**: For bone pathology tasks, the overall metrics indicate variability in performance (Fig. 3A). The model demonstrates fair diagnostic evidence accuracy on annotation with an average score of 3.58. However, it exhibits bad performance in diagnosis accuracy (2.56), evidence accuracy on texts (2.88), terminology accuracy (2.40) and multimodal information integration (1.93). Detailed question-level analysis (Fig. 3B) reveals overall low scores diagnosis accuracy and annotation evidence and integrating multimodal information in a majority of the cases. A significant variability in accuracy on evidence annotations is also identified.

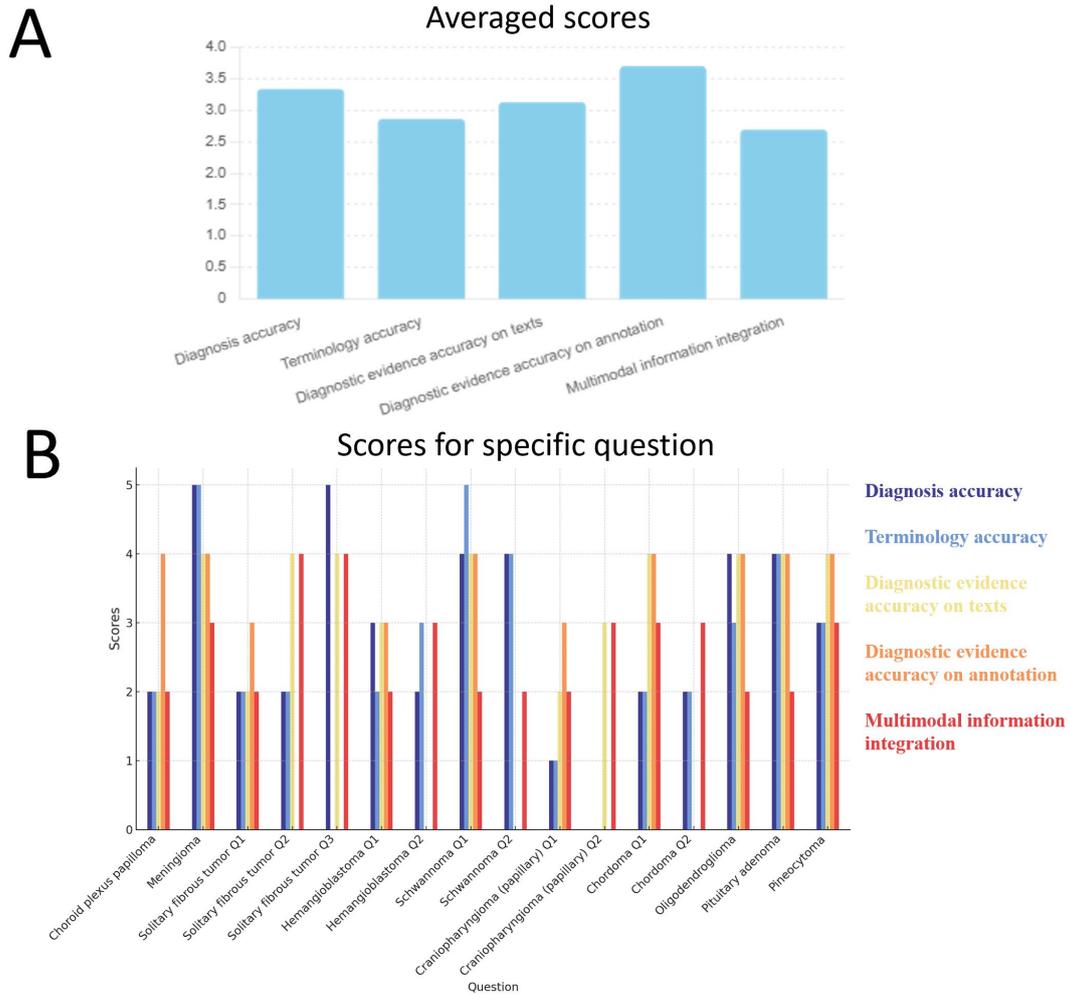

**Figure 4**. The detailed and averaged scores on the response of GPT for questions about CNS system. The original scores range from 1 to 5 and a zero-score indicates "non-applicable".

**CNS**: Further we evaluate ChatGPTs performance on CNS (central nervous system) pathology tasks. The overall metrics indicate a diverse range of performance (Fig. 4A). The model excels in diagnostic evidence accuracy on annotation, achieving an average score of 3.7, and demonstrates good diagnosis accuracy and diagnostic evidence accuracy on texts, with a score of 3.33 and 3.13 respectively. However, it shows bad performance in terminology accuracy (2.86) and multimodal information integration (2.69). These findings emphasize ChatGPTs strength in diagnostic annotation while identifying areas for improvement in specialized terminology and multimodal data integration for CNS pathology tasks. Fig. 4B, which provides a detailed breakdown of scores for individual questions, shows consistent levels of scores in diagnosis accuracy and annotation evidence. Also, there is a noticeable variability in terminology accuracy and challenges in integrating multimodal information.

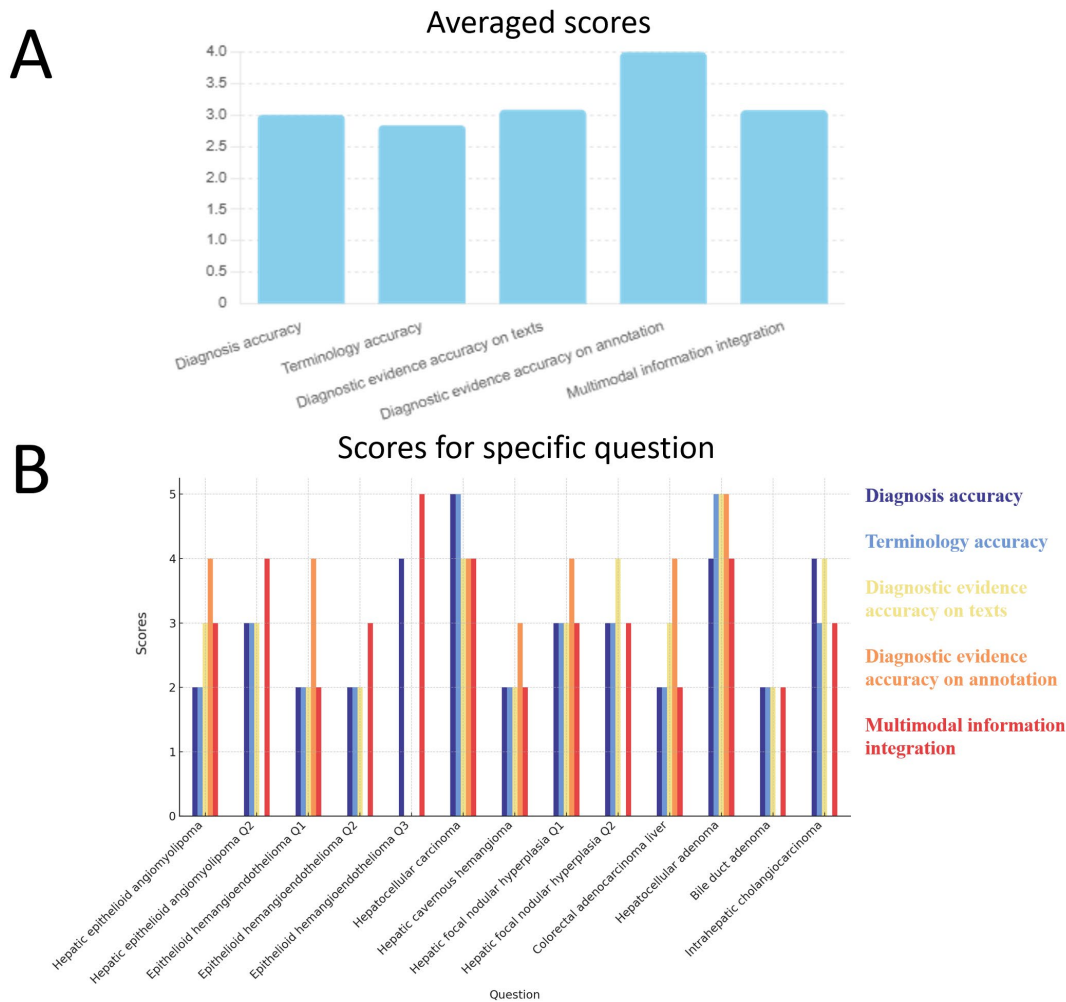

**Figure 5**. The detailed and averaged scores on the response of GPT for questions about liver system. The original scores range from 1 to 5 and a zero-score indicates "non-applicable".

**Liver** : The evaluation of ChatGPTs performance in liver pathology tasks reveals varied results across multiple metrics. Fig. 5A illustrates that ChatGPT performs best in diagnostic evidence accuracy on annotation with an average score of 4, followed by diagnosis evidence accuracy on texts at 3.08. The models performance s in diagnosis accuracy (3.00), terminology accuracy (2.86) and multimodal information integration (3.07) are fair. Detailed question-level analysis (Fig. 5B) highlights consistently low scores for conditions like "Hepatic cavernous hemangioma" in diagnosis accuracy and annotation evidence. The best performing cases are "Hepatocellular carcinoma" and "Hepatocellular adenoma".

*Case analysis with pathologists*

To offer insights into the specific deficits of ChatGPT model, we offer case studies on the question-answering about two patients, with detailed analyses on the responses.

**Case 1 with an ovarian Sertoli-Leydig cell tumor**: A 66-year-old female underwent oophorectomy due to a mass in the right ovary, and was diagnosed with an ovarian Sertoli-Leydig cell tumor. We provided an H&E image (200x magnification), which showed the tumor consisting of two components. One component had abundant, eosinophilic cytoplasm, was distributed in sheets, had vacuolated nuclei with small nucleoli visible. The other component interspersed within the first component, arranged in irregular gland-like structures, with abundant clear cytoplasm, deeply stained nuclei, and no obvious nucleoli. No clear mitotic figures were seen in the entire image.

In Fig. 6, based on the above H&E image, GPT provided the following description: 1. Cellular Atypia: The image shows obvious nuclear atypia, with varying nuclear sizes and staining intensities, indicating clear malignant characteristics. 2. High Cellular Density: The densely packed cells suggest a high cellular density, which is common in malignant tumors. 3. Disordered Structure: The tissue structure shows irregular arrangement, lacking normal tissue architecture, which is usually a sign of malignancy. 4. Cystic and Solid Components: The image presents a mixture of cystic and solid areas, which could be a feature of various types of ovarian tumors.

Considering the patient's age and clinical situation, these pathological features may be associated with the following types of ovarian tumors: 1. Ovarian Epithelial Cancer: The most common type of ovarian cancer, especially in older women. Epithelial tumors can present as a mixture of cystic and solid types. 2. Ovarian Serous Carcinoma: A specific type of epithelial cancer, often showing cellular atypia and high cellular density.

Further immunohistochemical testing is recommended to help determine the specific tumor type, including CA-125 and other relevant markers, which are crucial for accurate diagnosis and optimal treatment strategy selection. Additionally, considering the image characteristics, the patient's age, and the tumor's biological behavior, precise classification is essential for developing a treatment plan and prognosis assessment.

In this case, GPT's descriptions of cellular atypia and high cellular density were notably inaccurate and deviated significantly from the H&E image. The descriptions of disordered structure and cystic and solid components were more accurate, but the characteristic two components of this case were not described. Therefore, GPT was unable to effectively recognize and describe the tumor cells in this case, leading to significant discrepancies in the subsequent interpretation of IHC images and resulting in an inaccurate diagnosis.

We, based on GPT's description of the H&E image, provided two IHC images, AE1/AE3 and Calretinin. Based on the staining results of these two immunohistochemical markers, GPT provided the following conclusions: 1. AE1/AE3: This is a broad-spectrum cytokeratin antibody typically used to identify epithelial cells. Positive expression suggests an epithelial origin tumor, including various types of epithelial ovarian cancers. 2. Calretinin: This is a calcium-binding protein commonly used to diagnose tumors of mesothelial origin, such as mesotheliomas. In the diagnosis of ovarian tumors, positive expression of Calretinin helps distinguish mesothelial-origin tumors from other types.

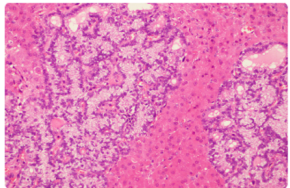
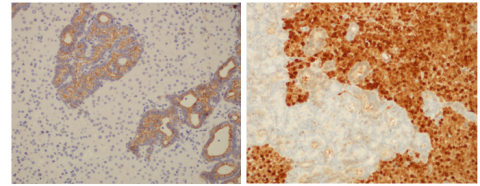

**Figure 6**. An exemplified case for GPT's response on a pathology images from a patient with an ovarian Sertoli-Leydig cell tumor.

Combining the positive expressions of these two markers: this combination suggests that the tumor may have a complex histological origin, including characteristics of both epithelial and mesothelial cells. The most likely diagnosis is ovarian mesothelioma or a borderline tumor, which may exhibit features of both epithelial and mesothelial cells. Further testing with additional immunomarkers (such as WT-1, CK7, CA-125, etc.) is recommended to further confirm the specific tumor type.

In fact, the positive signals of AE1/AE3 and Calretinin correspond to the two previously mentioned cellular components. The Leydig cell component expresses Calretinin but not AE1/AE3, while the Sertoli cell component shows the opposite pattern, expressing AE1/AE3 but not Calretinin. However, since GPT failed to recognize the two tumor cell components in this case, its interpretation of the IHC results was significantly flawed, leading to an incorrect diagnosis. Nevertheless, GPT's description of the uses of the two immunohistochemical antibodies was relatively accurate, demonstrating good knowledge of specific aspects, but its image recognition capabilities, especially precise recognition, still need improvement.

**Case 2 with colorectal cancer metastasis to the liver**: A 69-year-old male was found to have a liver mass for over a month and underwent liver tumor resection. The postoperative pathological diagnosis was colorectal cancer metastasis to the liver. We provided an H&E image (200x magnification), which showed tumor cells in the liver tissue arranged in irregular glandular and cribriform patterns, with an increased nucleus-to-cytoplasm ratio, vesicular nuclei, visible nucleoli, frequent mitotic figures, and a significant amount of necrosis within the glandular lumens.

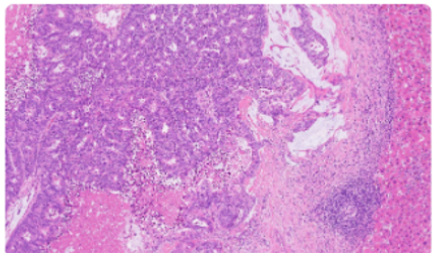

**Question:**
A 69-year-old male was found to have a liver mass for over a month. Based on this HE image, what do you think the diagnosis should be? Which immunohistochemical tests would you recommend to confirm the diagnosis?

**Answer:**
This HE-stained liver section shows some significant pathological features:

1. **Atypical Cell Proliferation**: In the center and lower part of the image, a large number of highly atypical cells can be observed. These cells are tightly packed with large, deeply stained nuclei, suggesting the possibility of a malignant tumor.
2. **Necrosis and Hemorrhage**: The lower left corner shows red blood cell accumulation and necrosis, indicating high tumor aggressiveness and vascular destruction.
3. **Fibrosis and Inflammatory Response**: On the right and upper parts of the image, there is noticeable fibrous tissue and inflammatory cell infiltration, possibly a reactive change to the tumor.

Considering the patient's clinical presentation and imaging findings, these features suggest the possibility of hepatocellular carcinoma (HCC) or cholangiocarcinoma. To further confirm the diagnosis, the following immunohistochemical tests are recommended:

- **Hepatocyte-specific antigen (Hep Par 1)**: To confirm a tumor of hepatocyte origin.
- **Alpha-fetoprotein (AFP)**: To help identify hepatocellular carcinoma.
- **Glypican-3 (GPC3)**: Highly specific for hepatocellular carcinoma.
- **Cholangiocyte markers (such as CK7 and CK19)**: To confirm a tumor of cholangiocyte origin.
- **Ki-67**: To assess the proliferative activity of the tumor; the higher the index, the more aggressive the tumor is likely to be.

**Figure 7**. An exemplified case for GPT's response on a pathology images from a patient with colorectal cancer metastasis to the liver.

As shown in Fig. 7, based on the above H&E image, GPT provided the following description and diagnostic opinions. This H&E-stained liver section shows some notable pathological features. 1. Atypical Cell Proliferation: In the center and lower left part of the image, there is a large number of highly atypical cells, tightly packed with large, deeply stained nuclei, suggesting the possibility of a malignant tumor. 2. Necrosis and Hemorrhage: The lower left corner shows red blood cell accumulation and necrosis, indicating high tumor aggressiveness and vascular destruction. 3. Fibrosis and Inflammatory Response: In the right and upper right part of the image, there is noticeable fibrous tissue and inflammatory cell infiltration, possibly a reactive change to the tumor.

Considering the patient's clinical presentation and imaging results, these features suggest the possibility of hepatocellular carcinoma (HCC) or cholangiocarcinoma. To further confirm the diagnosis, the following immunohistochemical tests are recommended: 1. Hepatocyte-specific antigen (Hep Par 1): To confirm a tumor of hepatocyte origin. 2. Alpha-fetoprotein (AFP): To help identify hepatocellular carcinoma. 3. Glypican-3 (GPC3): Highly specific for hepatocellular carcinoma. 4. Cholangiocyte markers (such as CK7 and CK19): To confirm a tumor of cholangiocyte origin. 5. Ki-67: To assess the proliferative activity of the tumor; the higher the index, the more aggressive the tumor is likely to be.

From the above description, it is evident that GPT's recognition and description of the tumor in this case were relatively accurate, likely due to extensive training on colorectal cancer tissue sections. The differential diagnosis provided by GPT includes relevant immunohistochemical suggestions. However, GPT failed to consider the possibility of colorectal cancer metastasis to the liver, focusing instead on primary liver tumors. This demonstrates GPT's limitation in flexibly applying learned knowledge, which is an issue that needs to be addressed in the future.

**Discussion**

This study presented a comprehensive evaluation of GPT-based models on pathology images. Using our self-constructed and clinical-oriented benchmark dataset, with 62 rounds of question answering covering four systems, we evaluated five dimensions of capacities of GPT on pathological diagnoses. Specifically, we identified significant pitfalls of GPT in interpreting immunohistochemistry results and diagnosing metastatic cancers with detailed case analyses. The results highlight both the potential and the limitations of such advanced AI system in pathological image analysis.

One of the key findings is the generalist multimodal AI models capacity in understanding and describing complex visual patterns in pathology images remains limited. Specifically, the study identifies several specific areas where improvements are necessary. First, GPT models are currently unable to accurately identify tumor regions on IHC images, which hinders their ability to correctly interpret these critical diagnostic tools. A training on well-annotated IHC images or paired H&E and IHC images with a sufficient data scale may solve this issue. Second, GPT exhibits restricted consideration of metastatic diseases, demonstrating a need for more sophisticated algorithms capable of broader diagnostic reasoning. Additionally, GPT models struggle to correctly integrate imaging information necessary for clinical diagnosis when presented with complex cases that require such integration. This highlights a significant gap in their current capabilities, where the nuanced interpretation of combined imaging and pathology data is crucial. The accuracy of translating pathological diagnostic terms also needs improvement, with frequent translation errors potentially leading to misdiagnoses. For diseases with overlapping morphology, GPT models find it challenging to provide accurate diagnoses based on a single H&E image. Future effort using whole slide imaging could potentially enhance diagnostic accuracy significantly. This indicates a clear direction for future research, emphasizing the need for comprehensive datasets and advanced imaging techniques to train these models.

Despite these challenges, the evaluation reveals several advantages of GPT in pathological diagnosis. GPT models possess accurate and wide-ranging knowledge of pathology, which can be advantageous for educational purposes and answering question banks. Moreover, these models can suggest relevant clinical treatment methods based on the pathological information provided by images and demonstrate a strong understanding of the application and staining locations of immunohistochemical antibodies. Importantly, the more comprehensive the information provided (including H&E images, IHC results, molecular testing results, and clinical history), the higher the diagnostic accuracy GPT can achieve. Finally yet interestingly, despite the restricted diagnosis performance, the annotations as diagnostic evidence are much satisfactory in all tested systems. This ability from GPT model can partially aid pathologists in diagnosis and treatment planning.

We understand the restricted performance of GPT from several attributions. The variability and complexity of pathology images, coupled with the limited availability of high-quality annotated datasets, pose significant challenges for AI model training and evaluation [7, 15]. These factors can affect the generalist models accuracy and generalizability across different pathological conditions and imaging modalities [16, 17]. Given the emerging pathology specialist models [6, 18, 19, 20], a combination between generalist and specialist models could be an alternative approach to achieve realistic applications with advanced AI [21]. Additionally, continuous updates and validation of AI models are necessary to keep pace with advancements in medical imaging and pathology practices. Correspondingly, future research should focus on expanding and diversifying pathology image datasets, improving model-training techniques, and developing robust validation frameworks. Collaborative efforts between AI researchers, pathologists, and clinical practitioners are essential to address these challenges and unlock the full potential of AI in pathology. Furthermore, the integration of AI models into clinical workflows requires careful consideration of ethical, legal, and practical aspects. Ensuring the transparency, interpretability, and reliability of AI-generated reports is crucial for gaining the trust of healthcare professionals and patients [22, 23].

In conclusion, benchmarking GPT-based models on pathology images demonstrates a promising step towards integrating advanced AI technologies into medical imaging. By enhancing diagnostic accuracy and efficiency, these models have the potential to revolutionize pathology and contribute to the broader goal of personalized medicine. Continued research and development are imperative to overcome current limitations and ensure the successful implementation of AI in clinical settings.

**Consent for publications**
Not applicable.

**Availability of data and materials**
The data will be made publically available via Github upon acceptance of the paper.

**Acknowledgments**
This study is supported in part by Shanghai Artificial Intelligence Laboratory.

**Conflict of interest**
All authors declare no potential conflict of interests.

**Author's contributions**
Conceptualization, ML, JW, SZ, and ZW; Methodology, ML, JW; Software, FY and HL; Investigation, ML, JW, and WW; Resources, SZ and ZW; Data curation, JW, WW, and ZW; Writing—original draft preparation, ML, JW; Writing—review and editing, ML, JW, FY, SZ and ZW; Visualization, ML and FY; Supervision, SZ and ZW; Funding acquisition, SZ and ZW; All authors have read and agreed to the published version of the manuscript. All authors contributed to the article and approved the submitted version.